\documentclass[twocolumn,showpacs,preprintnumbers,amsmath,amssymb,floatfix]{revtex4}

\usepackage{graphicx}
\usepackage{dcolumn}
\usepackage{bm}

\begin{document}


\title{Generating Schr\"{o}dinger-cat states in momentum and internal-state space from
Bose-Einstein condensates with repulsive interactions}
\author{J.\ Higbie and D.\ M.\ Stamper-Kurn}

\affiliation{Department of Physics, University of California,
Berkeley CA  94720
}%

\date{\today}

\begin{abstract}
Resonant Raman coupling between internal levels induced by
continuous illumination of non-collinear laser beams can create
double-well momentum-space potentials for multi-level
``periodically-dressed'' atoms.  We develop an approximate
many-body formalism for a weakly interacting, trapped
periodically-dressed Bose gas which illustrates how a tunable
exchange interaction yields correlated many-body ground states. In
contrast to the case of a position-space double well, the ground
state of stable periodically-dressed Bose gases with repulsive
interactions tends toward a Schr\"{o}dinger cat state in the
regime where interactions dominate the momentum-space tunnelling
induced by the external trapping potential.  The dependence of the
momentum-space tunnelling and exchange interaction on experimental
parameters is derived.  We discuss how real-time control of
experimental parameters can be used to create Schr\"{o}dinger cat
states either between  momentum or internal states, and how these
states could be dynamically controlled towards highly sensitive
interferometry and frequency metrology.
\end{abstract}

\maketitle{}

\newcommand{\hami}{\hat{\mathcal{H}}}
\newcommand{\hamiint}{\hat{\mathcal{H}}_{\mbox{\small int}}}
\newcommand{\twocolvec}[2]{\left( \begin{array}{c} #1 \\ #2
\end{array} \right)}
\newcommand{\tworowvec}[2]{\left( \begin{array}{c c} #1 & #2
\end{array} \right)}
\newcommand{\twotwomatrix}[4]{\left( \begin{array}{c c} #1 & #2 \\
#3 & #4
\end{array} \right)}


Following our mastery over the internal and external states of
individual atoms, the scientific frontier advances to the full
control over the quantum states of many-body systems. Entangled
states, in which the constituents of a many-body system display
non-classical correlations, play a key role in our developing
understanding of quantum information and decoherence, and may find
practical use in quantum communication \cite{benn93tele}, quantum
computing, and high-precision metrology
\cite{cave81,wine92squeeze,boll96,bouy97}. Various techniques are
being developed that generate entanglement deterministically in,
for example, spin-squeezed atomic ensembles \cite{juls01}, trapped
ions \cite{sack00four}, the electromagnetic field
\cite{brun96meter}, and superconducting circuits
\cite{pash03,well03}.

Ultracold neutral atoms offer a promising route to generating
highly entangled many body states. Schemes for generating such
entanglement rely on interatomic interactions which provide
``non-linear'' elements as seen from the viewpoint of
single-particle dynamics.  Such non-linear terms are provided
naturally through binary collisions between ground-state atoms, as
utilized controllably by Mandel \emph{et al.}\ \cite{mand03cont},
and can be further enhanced through molecular resonances or by the
assistance of near-resonant photons
\cite{andr02corr,sore02cavity}.

A paradigmatic system in which interatomic interactions induce
entanglement is a collection of ultracold interacting bosons in a
parity-symmetric double-well potential
\cite{imam97,cira98super,java99split,spek99}. The lowest-energy
single-particle states are the parity-even ground state
$|\psi_0\rangle$ and parity-odd first excited state
$|\psi_1\rangle$, separated in energy by the tunnel splitting $J$.
If tunnelling and interactions are weak with respect to the energy
spacings to other single-particle states, the many-body system may
be described by the  two-mode Hamiltonian
\begin{equation}
\label{eq:position_double_well} \mathcal{H} = -\frac{J}{2} \left(
\hat{c}^\dagger_R \hat{c}_L + \hat{c}^\dagger_L \hat{c}_R \right)
- U \hat{N}_R \hat{N}_L
\end{equation}
whence terms dependent only on the total number of atoms in the
system are omitted.  Here $\hat{c}_R$ and $\hat{c}_L$ denote
annihilation operators for particles in the right- ($|R\rangle =
[|\psi_0\rangle + |\psi_1\rangle]/\sqrt{2}$) or left- ($|L\rangle
= [|\psi_0\rangle - |\psi_1\rangle]/\sqrt{2}$) well states,
respectively. The number operators $\hat{N}_R = \hat{c}^\dagger_R
\hat{c}_R$ and $\hat{N}_L = \hat{c}^\dagger_L \hat{c}_L$ count
particles in the right or left wells, respectively, of the
double-well system.  The parameter $U$ gives the energy due to
short-range interactions of a pair of atoms located in the same
well.

This simple Hamiltonian leads to highly-correlated many-body
ground states through the interplay of tunnelling and interaction.
One finds three limiting behaviors.  If the tunneling rate
dominates, the many-body ground-state of $N$ bosons is driven to
the factorized, uncorrelated state in which all $N$ atoms are
identically in the single-particle ground state $|\psi_0\rangle$.
In the limit $N |U|/J \gg 1$, the interaction energy dominates.
For repulsive interactions ($U
> 0$), a many-body state divides itself evenly between the two
potential minima, generating the state $|\Psi\rangle \propto
(\hat{c}^\dagger_R)^{N/2} (\hat{c}^\dagger_L)^{N/2} |0\rangle$,
where $|0\rangle$ is the vacuum state.  Experimental evidence for
such a ``number-squeezed'' state has been obtained by Orzel
\emph{et al.}\ in a many-well potential \cite{orze01squeeze}.
Similar physics is responsible for the observed Mott-insulator
phase  in a three-dimensional periodic potential
\cite{grei02mott}.

For attractive interactions ($U < 0$), the ground-state is quite
spectacular: a Schr\"{o}dinger-cat state $|\Psi\rangle \propto
[(\hat{c}^\dagger_R)^{N} + (\hat{c}^\dagger_L)^{N}] |0\rangle$
formed as a superposition of states in which \emph{all} atoms are
found in one of the two wells.  The generation of Schr\"{o}dinger
cat states is a tantalizing goal; at present, the largest atomic
Schr\"{o}dinger-cat states contain just four atoms
\cite{sack00four}. However, accomplishing this goal by imposing a
double-well potential on a collection of bosons with attractive
interactions is a daunting task: such Bose-Einstein condensates
are unstable to collapse \cite{sack99collapse}, implying that the
number of bosons placed into the superposition state will be
small; moreover, exceptional spatial control over trapping
potentials and over the interaction strength is required.

The above discussion pertains to atoms for which the
\emph{position-space potential is a double well}, while the
kinetic energy term $p^2/2 m$ may be regarded as a
\emph{momentum-space harmonic ``potential.''}  In this work, we
point out that an analogous many-body system can be crafted, in
which the roles of position and momentum are interchanged.  As
discussed in previous work \cite{higb02,stam03anis,mont03}, the
dispersion relation of multi-level atoms placed in a
spatially-periodic coupling between internal states can take the
form of a \emph{momentum-space double-well ``potential.''}
Addition of \emph{an harmonic position-space trapping potential}
produces a double-well system dual to that discussed above.
Assessing the roles of interparticle interactions in this
situation, we find that the many-body ground states for bosonic
atoms in such a system are highly-entangled in both momentum- and
internal-state space.  In opposition to the situation of bosons in
position-space double wells, maximally-entangled states may be
generated in the case of \emph{repulsive} interactions, allowing
the creation of such states starting with large, stable
Bose-Einstein condensates while obviating the need for exacting
control over potentials with extremely small spatial dimensions.

The roles of repulsive or attractive interactions in determining
the behaviour of interacting bosons in momentum-space double-well
potentials are reversed from those in the case of position-space
potentials.   As discussed below, this reversal is due to an
exchange term which arises in the evaluation of the interaction
energy of a Bose gas that occupies several distinct momentum
states. For repulsive interactions, for example, this exchange
term disfavors a macroscopic occupation of more than one
single-particle wavefunction (the so-called fractionation of a
condensate as discussed by Leggett \cite{legg00rmp} and others),
thus favoring a Schr\"{o}dinger-cat superposition.  We find that
the strength of the exchange term can be dynamically varied by
varying parameters of the periodic coupling field which generates
the double-well potential.

Following a derivation of the momentum-space double-well potential
(Sec.\ \ref{sec:origin}), we develop a two-mode approximate
treatment of this system which provides expressions for the
interaction and tunnelling energies and thereby clarifies their
dependence on experimental parameters (Sec.\ \ref{sec:twomode}).
Similar work by Arecchi and Montina \cite{mont03} treats this
system using the Gross-Pitaevskii equation as the starting point,
a numerical scheme which identifies the onset of correlated
many-body ground states.  Their work appears to reproduce our
analytic approach in the limit of weak interactions, while the
validity of their approach (or, indeed, any extant treatment) for
stronger interactions is difficult to establish. Furthermore, in
Section \ref{sec:heisenberg}, we show that maximal entanglement
can be generated purely between internal states or between
momentum states, offering a route to Heisenberg-limited atomic
clocks \cite{wine92squeeze} or atomic interferometry
\cite{bouy97}, respectively.

\section{Origin of the momentum-space potential \label{sec:origin}}

\begin{figure}
    \begin{center}
    \center{\includegraphics[width=3 in]{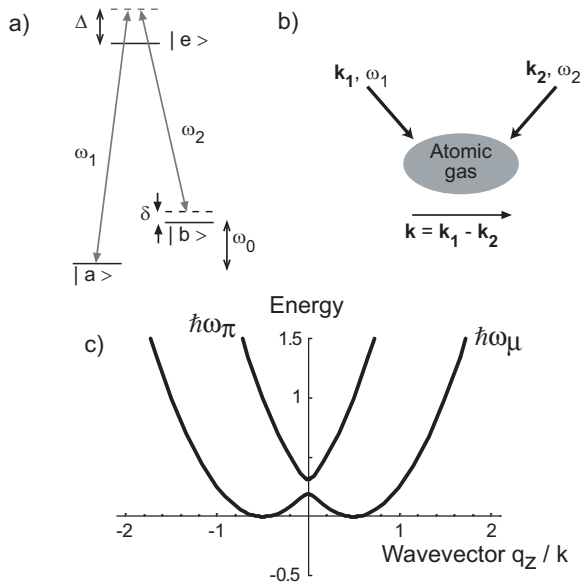}}
     \caption{Creating a double-well momentum-space potential with Raman excitation.  (a) Laser beams of
     frequency $\omega_1$ and $\omega_2$ may induce Raman
     transitions between internal states $|A\rangle$ and
     $|B\rangle$.  $\delta = (\omega_1 - \omega_2) - \omega_0$ is the detuning from the Raman
     resonance.  (b) Such a Raman transition imparts a momentum transfer of
     $\hbar {\bf{k}} = \hbar ({\bf{k}}_1 - {\bf{k}}_2)$, where ${\bf{k}}_1$ and
     ${\bf{k}}_2$ are the wavevectors of the Raman coupling
     lasers.  (c) Atoms exposed to continuous Raman excitation can be described by a two-branch dispersion relation.  Energies (scaled by $E_k$) for the
     lower ($\hbar \omega_\mu$) and upper ($\hbar \omega_\pi$) dressed states are shown for wavevectors (scaled by $k$) in the direction of the Raman momentum transfer.
     In the case of exact Raman resonance ($\delta = 0$) and small Raman Rabi frequency (shown for $\hbar \Omega/E_k = 1/8$),
     the lower dispersion relation takes the form
     of a parity-symmetry double-well potential.
     }
     \label{fig:scheme}
    \end{center}
\end{figure}

We consider bosonic atoms of mass $m$ with two internal states,
$|A\rangle$ and $|B\rangle$, at energies $\hbar \omega_A$ and
$\hbar \omega_B$, respectively. These atoms are exposed to laser
fields of frequencies $\omega_1$ and $\omega_2$ ($\omega =
\omega_1 - \omega_2$) and wavevectors ${\bf{k}}_1$ and
${\bf{k}}_2$ which induce Raman transitions between the two
internal states (see Fig.\ \ref{fig:scheme}a,b).  Thus, a Raman
process from state $|A\rangle$ to $|B\rangle$ imparts a momentum
of $\hbar {\bf{k}} = \hbar ({\bf{k}}_1 - {\bf{k}}_2)$ and kinetic
energy of $\hbar \delta = \hbar \omega - \hbar(\omega_B -
\omega_A)$.  The continuous Raman coupling can be regarded as a
spatially-periodic coupling field between the internal states of
the atoms, represented by an off-diagonal potential $V_{R} = \hbar
\Omega/2 \left(e^{-i ({\bf{k}}\cdot {\bf{r}} - \omega t)}
|A\rangle\langle B| + e^{i ({\bf{k}}\cdot {\bf{r}} - \omega t)}
|B\rangle\langle A| \right)$. Here $\Omega$ is the two-photon Rabi
frequency, taken to be real, which is determined by dipole matrix
elements, the detuning from intermediate resonances, and by the
laser intensities.

It is convenient to analyze this constantly-driven system in terms
of ``periodically-dressed'' states, which are coherent
superpositions of both internal and external (momentum) states. As
developed in Refs.\ \cite{higb02,stam03anis}, this treatment
reveals marked anisoptropy and tunability in the superfluid
characteristics of a Bose-Einstein condensate formed of such a
dressed-state gas. Given a two-component spinor wavefunction
$\vec{\psi}({\bf{r}}) = ( \psi_A({\bf{r}}), \psi_B({\bf{r}}) )$ to
describe single-atom states in the $\{|A\rangle, |B\rangle\}$
basis, we transform to a frame which is \emph{co-rotating} and
\emph{co-moving} with the driving laser fields, yielding a spinor
wavefunction $\vec{\widetilde{{\psi}}}({\bf{r}})$ with components
$\widetilde{\psi}_{A,B}({\bf{r}}) = \exp(\pm i ({\bf{k}} \cdot
{\bf{r}} - \omega t)/2) \psi_{A,B}({\bf{r}})$. This transformation
yields a Schrodinger equation with two deBroglie wave solutions at
each momentum $\hbar {\bf{q}}$, which we call the the
periodically-dressed states. The two-branch dispersion relation of
the periodically-dressed atoms has the form (using $\pi$ for plus
and $mu$ for minus)
\begin{equation}
 \hbar \omega_{\pi,\mu}({\bf{q}}) = {q}^2 + \frac{1}{4} \pm \frac{1}{2} \sqrt{
\left( 2 {\bf{q}} \cdot {{\bf{k}}} - {\delta}\right)^2 +
{\Omega}^2}
\end{equation}
where energies and wavevectors are scaled by the Raman recoil
energy $E_k = \hbar^2 k^2/2 m$ and wavevector $k$, respectively.

Under the conditions of exact Raman resonance ($\delta = 0$) and
sufficiently small Rabi frequency ${\Omega}$, a degeneracy of
momentum ground states occurs as the lower dispersion relation
takes the form of a double-well potential, with minima at $\pm
{\bf{Q}}/2$ (${\bf{Q}} \simeq {{\bf{k}}}$) (Fig.\
\ref{fig:scheme}c). This potential can be quickly tuned by
modifying the laser based parameters. The detuning from Raman
resonance $\delta$ breaks the ground-state degeneracy, favoring
the right- or left-well states. The spacing between the potential
wells is controlled by the Raman momentum transfer $\hbar
{\bf{k}}$, which can be varied by reorienting the laser beams.
Finally, the Rabi frequency $\Omega$ changes both the height of
the barrier between the wells, as well as the internal-state
character of states on either side of the well \cite{stam03anis}.

In the case of a position-space double well, the kinetic energy of
the atoms forbids a complete localization of atoms in either of
the wells, thus introducing tunnelling.  Similarly, in our case of
the momentum-space double-well, a tunnelling between well-defined
momentum states can be induced by adding a spatially-dependent
term to the Hamiltonian.  We consider adding an internal-state
independent position-space trapping potential of the form
$V({\bf{r}}) =  m \omega_t^2 r^2 / 2$
\cite{radial_symmetry_footnote}. The Hamiltonian for this system
may be written in the basis of the periodically-dressed states
introduced earlier, using the expansion
\begin{equation}
\vec{\widetilde{\psi}}({\bf{r}}, t)  =
 \int \! \frac{d^3 {\bf{q}}}{(2 \pi)^{3/2}}
\mathrm{R}({\bf{q}}) \left( \begin{array}{c} \pi({\bf{q}}) \\
\mu({\bf{q}}) \end{array} \right) \, e^{i {\bf{q}} \cdot {\bf{r}}}
\end{equation}
where $\mathrm{R}({\bf{q}}) = e^{-i \sigma_y \theta({\bf{q}})/2}$
where $\sigma_y$ is a Pauli matrix and the mixing angle
$\theta({\bf{q}})$ is defined by the relation $\cot
\theta({\bf{q}}) = (\delta - \hbar {\bf{q}} \cdot {\bf{k}}/m) /
\Omega$.  The wavefunction $\vec{\phi}({\bf{q}}) = (\pi({\bf{q}}),
\mu({\bf{q}}) )$ in the space of periodically-dressed momentum
eigenstates obeys a Schr\"{o}dinger equation with Hamiltonian
\begin{equation}
\hami_{PD} = \hbar \left(\begin{array}{c c} {\omega}_+({\bf{q}}) &  0 \\
0 & {\omega}_-({\bf{q}})
\end{array} \right) - \frac{1}{M^2} \mathrm{R}^\dagger({{\bf{q}}})
\nabla_{{q}}^2 \mathrm{R}({{\bf{q}}})
\label{eq:singleparticlehami}
\end{equation}
The position-space trapping potential is seen in momentum space as
a kinetic-energy-like term (involving $\nabla^2_q$), which
accounts also for the variations with ${\bf{q}}$ of the
periodically-dressed eigenstate basis. The relevance of this
kinetic-energy-like term is measured by the dimensionless
effective mass parameter $M$, which is related to the Lamb-Dicke
parameter $\eta = \sqrt{\hbar k^2 / 2 m \omega_t}$ as $M = 2
\eta^2$. For weak spatial confinement ($M \gg 1$), the low-energy
single-particle states are restricted primarily to the lower
periodically-dressed states, with the two lowest states split by a
small tunnelling energy $J$. For strong confinement ($M \ll 1$),
the single-particle states are admixtures of upper- and
lower-dispersion-relation periodically-dressed states, and a
simple double-well treatment is no longer adequate.

To introduce interatomic interactions, we make use of the field
operator $\hat{\phi}({\bf{q}}) = \left(\hat{\pi}({\bf{q}}),
\hat{\mu}({\bf{q}}) \right)$ the components of which annihilate
particles in the lower ($\mu$) or upper ($\pi$)
periodically-dressed states. These operators obey Bose commutation
relations.  We consider low-energy, binary, elastic collisions
with a state-independent scattering length $a$. The interaction
Hamiltonian then takes the form $\hamiint = \frac{g}{2} \left(
\int d^3 {\bf{q}} \, \hat{n}({\bf{q}}) \hat{n}(-{\bf{q}})\right)$,
neglecting terms dependent only on the total number of atoms $N$,
where $\hat{n}({\bf{q}})$ is the spatial Fourier transform of the
density operator, and $g = 8 \pi k a$ is the properly scaled
interaction parameter. The density operator is given as
\cite{higb02,stam03anis}
\begin{equation}
\hat{n}({\bf{q}}) = \int \! \frac{d^3 {\bf{\kappa}}}{(2
\pi)^{3/2}} \hat{\phi}^{\dagger}({\bf{\kappa}} +
\frac{{\bf{q}}}{2}) \mathrm{R}^\dagger({\bf{\kappa}} +
\frac{{\bf{q}}}{2}) \mathrm{R}({\bf{\kappa}} - \frac{{\bf{q}}}{2})
\hat{\phi}({\bf{\kappa}} - \frac{{\bf{q}}}{2})
\end{equation}

\section{The two mode approximation \label{sec:twomode}}

To simplify our treatment, let us consider only the situation in
which the two lowest energy eigenvalues (in the absence of
interactions) are well separated from the remaining energies, and
thus a two-mode description of the many-body system is adequate.
The two lowest-energy single-particle states have wavefunctions
$\vec{\phi}_0({\bf{q}})$ and $\vec{\phi}_1({\bf{q}})$, mode
operators $\hat{c}_0$ and $\hat{c}_1$, and energy splitting $J$.
We define the right- and left-well states as
$\vec{\phi}_{R,L}({\bf{q}}) = \left( \vec{\phi}_0({\bf{q}}) \pm
\vec{\phi}_1({\bf{q}})\right) / \sqrt{2}$, and the mode operators
as  $\hat{c}_{R,L} = (\hat{c}_0 \pm \hat{c}_1)/ \sqrt{2}$.  Under
this approximation, we can evaluate the density operator
$\hat{n}({\bf{q}})$  by expanding the field operators
$\hat{\mu}({\bf{q}})$ and $\hat{\pi}({\bf{q}})$ in the basis of
energy eigenstates, and then truncating the expansion after the
first two states. We then express the density operator as
$\hat{n}({\bf{q}}) \simeq \sum_{i,j} \mathcal{N}_{i j}({\bf{q}})
\hat{c}^\dagger_i \hat{c}_j $ with
\begin{equation} \label{eq:nfunctions}
\mathcal{N}({\bf{q}})_{i j} = \int \!  \frac{d^3 {\bf{\kappa}}}{(2
\pi)^{3/2}} \vec{\phi}_i^\dagger ({\bf{\kappa}} +
\frac{{\bf{q}}}{2}) \mathrm{R}({\bf{\kappa}} +
 \frac{{\bf{q}}}{2}) \mathrm{R} ({\bf{\kappa}} -  \frac{{\bf{q}}}{2})
\vec{\phi}_j({\bf{\kappa}} + \frac{{\bf{q}}}{2})
\end{equation}
with indices $i, j \in \{R, L\}$ denoting either the right or left
states.  Accounting for properties of the right- and left-well
states under parity \cite{symmetry_footnote}, we obtain
\begin{equation} \hat{n}({\bf{q}}) \simeq \mathcal{N}_{R
R}({\bf{q}}) N + \mathcal{N}_{R L}({\bf{q}}) \hat{c}^\dagger_R
\hat{c}_L + \mathcal{N}_{R L}(-{\bf{q}}) \hat{c}^\dagger_L
\hat{c}_R
\end{equation}
where $N$ is the total number of atoms in the system.  Dropping
terms that depend only on $N$, one thus finds
\begin{eqnarray}
\hamiint & \simeq&  -U \hat{N}_R \hat{N}_L + \frac{J_1}{2} \left(
\hat{c}^\dagger_R \hat{c}_L
 + \hat{c}^\dagger_L
\hat{c}_R\right) \nonumber \\
& & + J_2 \left(\hat{c}^\dagger_R \hat{c}^\dagger_R \hat{c}_L
\hat{c}_L + \hat{c}^\dagger_L \hat{c}^\dagger_L \hat{c}_R
\hat{c}_R \right)  \label{eq:hint_two_mode}
\end{eqnarray}
where the energies $U$, $J_1$ and $J_2$ are given as
\begin{eqnarray}
U & = & - g \int d^3 {\bf{q}} \, \mathcal{N}^2_{R L} \\
J_1 & = &  2 N g  \int d^3 {\bf{q}}  \mathcal{N}_{R R}({\bf{q}})
\mathcal{N}_{R L}({\bf{q}}) \\
J_2 & = & \frac{g}{2} \int d^3 {\bf{q}} \,  \mathcal{N}_{R
L}({\bf{q}}) \mathcal{N}_{R L}(-{\bf{q}})
\end{eqnarray}

Before delving further into the implications of the interaction
Hamiltonian given above, let us consider the weak confinement
(large $M$) limit at which  the right- and left-well states
contain no population in the upper ($\pi$) dressed states, are
Gaussian functions well-localized (in ${\bf{q}}$ space) at the
right and left potential minima at $\pm {\bf{Q}}/2$, with rms
widths $\sigma_q \simeq (2 M)^{-1/2}$, and have average density
per particle $\langle n \rangle = \sigma_q^3 / \pi^{3/2}$. In
calculating the integrals of Eq.\ \ref{eq:nfunctions}, we may thus
use local values of the rotation matrices $\mathrm{R}({\bf{q}})$
at $\pm {\bf{Q}}/2$, as appropriate.  We then find
\begin{eqnarray} \label{eq:startscalings}
U & = & -g  \langle n\rangle \sin^2 \theta\left(\frac{{\bf{Q}}}{2}\right) \\
J_1 & = & 2 N g  \langle n\rangle \sin
\theta\left(\frac{{\bf{Q}}}{2}\right) \times e^{- \frac{Q^2}{16
\sigma_q^2}} \\
J_2 & = & \frac{g}{2} \langle n\rangle \sin^2
\theta\left(\frac{{\bf{Q}}}{2}\right) \times e^{- \frac{Q^2}{4
\sigma_q^2}} \label{eq:endscalings}
\end{eqnarray}
These approximations can be further simplified by setting
${\bf{Q}} = {{\bf{k}}}$, and thus $\sin\theta({\bf{Q}}/2)) \simeq
\sqrt{\Omega^2 / (1 + \Omega^2)}$.

Let us presume that the interaction-mediated tunnelling terms
$J_1$ and $J_2$ are negligibly small.  We thus recover the
many-body Hamiltonian of Eq.\ \ref{eq:position_double_well} which
had applied to the \emph{position-space double well}.  However,
examining the interaction energy $U$, we now find that \emph{the
roles of attractive and repulsive interactions are interchanged}
between the position-space and momentum-space double well
treatments: for the case of the momentum-space double-well
potential, repulsive interactions ($g>0$) yield $U < 0$ and thus
are the source of Schr\"{o}dinger-cat ground states, while
attractive interactions ($g<0$) yield $U > 0$ and thus are the
source of number-squeezed ground states.  This opens the door to
the production of Schr\"{o}dinger-cat states for large, robust
Bose-Einstein condensates with repulsive interactions through the
use of binary collisions as a non-linear coupling.

This reversal of roles for repulsive and attractive interactions
in momentum space can be understood in the context of a
single-component Bose-Einstein condensate in a harmonic potential.
Repulsive interactions lead the ground-state condensate
wavefunction to become larger spatially, thus causing  the
momentum-space wavefunction to contract -- appearing as an
attractive interaction in momentum-space.  Equivalently, one may
note that in a Hartree approximation to the many-body state of a
Bose-Einstein condensate, the interaction energy is evaluated as
an interaction energy density proportional to the square of the
density.  The density of a Bose gas occupying two distinct
momentum states would be spatially modulated by the interference
between the momentum states, and, therefore, its interaction
energy $\propto g n^2$ would be increased (decreased) for the case
of repulsive (attractive) interactions.  Thus we find repulsive
interactions leading to Schr\"{o}dinger-cat states, a
superposition of states in which only one single-particle state is
macroscopically occupied, while attractive interactions would lead
to a ``fractionation'' of the condensate between distinct momentum
states.

In the system we have described here, the interaction energy $U$
is a \emph{tunable exchange term} which, in the case of repulsive
interactions, suppresses the superposition of atoms in two
distinct momentum states and thereby favors the
maximally-entangled state.  The exchange term in the interaction
Hamiltonian arises from the presence of atoms in identical
internal states, but different momentum states.  By adjusting the
strength of the Rabi coupling $\Omega$, the admixture of internal
states $|A\rangle$ and $|B\rangle$ in the right and left well
states is varied.  For small $\Omega$, atoms in the right-well
state are almost purely in the $|A\rangle$ state, and atoms in the
left-well state are almost purely in the $|B\rangle$ state.  Thus,
the strength of the exchange term is suppressed ($\sin^2
\theta({\bf{Q}}/2) \ll 1$), and the product state $|\Psi\rangle
\propto (\hat{c}_R + \hat{c}_L)^N|0\rangle$ remains energetically
favored. For larger $\Omega$, the right and left-well states both
contain admixtures of the two internal states; e.g.\ the
right-well state contains atoms in state $|A\rangle$ nearly at
rest, while the left-well state contains atoms in state
$|A\rangle$ at momentum $\sim -\hbar {\bf{k}}$.  Significant
exchange terms now suppress the product state in favor of the
maximally-entangled state.

The dependence of the many-body ground state on the ratio $U/J$
and the atom number $N$ has been worked out by various authors
\cite{imam97,spek99,java99split} for the position-space
double-well potential. As we have obtained a Hamiltonian identical
in form to the position-space treatment, these predictions would
apply directly to our scheme. Significant deviations from the
factorized, non-interacting atom solution begin at $|U/J| \gtrsim
1/N$; for repulsive ($U > 0$) or attractive ($U < 0$) interactions
this would shift population either away from or towards the state
$|N_L = N/2, N_R = N/2\rangle$, respectively. Perturbation
analysis shows that the many-body state will be well approximated
by the Schr\"{o}dinger-cat state $(|N, 0\rangle + |0,
N\rangle)/\sqrt{2}$ when $U/J > 1/\sqrt{N}$.

Now returning to Eq.\ \ref{eq:hint_two_mode}, we see that two
extra terms appear in $\hamiint$ which do not conserve the number
of particles in the right- and left-well states, representing a
form of interaction-mediated tunnelling. The first, involving
$J^{(1)}$, modifies the tunnelling energy between the right and
left wells. The effect of this term is to reduce the tunnelling
rate for repulsive interactions, and increase it for attractive
interactions.  This can be understood by noting that repulsive
interactions tend to raise the energy of the single-particle
ground state more than the first excited state since the ground
state has a higher density. Our previous analysis of the two-mode
approximation accommodates this term by a redefinition of $J$ as
$J \rightarrow J - J^{(1)}$, leading to no further complications.
The second tunnelling term, involving $J^{(2)}$, describes
collisions which can redistribute two atoms from the right well to
the left well, and vice versa. While this term modifies the
conclusions of our simple treatment, we have seen that the
magnitudes of both $J^{(1)}$ and $J^{(2)}$ are exponentially
suppressed with respect to $U$, and thus the dominant role of
interactions is to create the aforementioned many-body ground
states.  These terms will, however, play an important role in
determining the strength of interactions at which many-body
correlations begin to become evident; the scaling of $J$
(independent of $g$), $J^{(1)}$, $J^{(2)}$ and $U$  with the
control parameters ($\Omega$, $g$, $M$, etc.) are different and
thereby provide more tunability to the system.

We performed numerical calculations of the lowest-energy
single-particle states to verify the simple scaling behaviour of
various terms in the Hamiltonian described above.  The
single-particle Hamiltonian of Eq.\ \ref{eq:singleparticlehami} is
separable in the three Cartesian coordinates defined so that one
axis (say $\hat{z}$) lies along the direction of the Raman
momentum transfer ${\bf{k}}$.  We then have the lowest energy
eigenstates as $\vec{\phi}_{0,1}({\bf{q}}) = \vec{\phi}_{0,1}(q_z)
\times \exp(-(q_x^2 + q_y^2)/4 \sigma_q^2) / \sqrt{2 \pi
\sigma_q^2} $, a product of harmonic-oscillator ground states in
the  $\hat{x}$ and $\hat{y}$ directions and normalized solutions
$\vec{\phi}_{0,1}(q_z)$ to the one-dimensional, two-component
Schr\"{o}dinger equation derived from Eq.\
\ref{eq:singleparticlehami}.

\begin{figure}
    \begin{center}
    \center{\includegraphics[width=3 in]{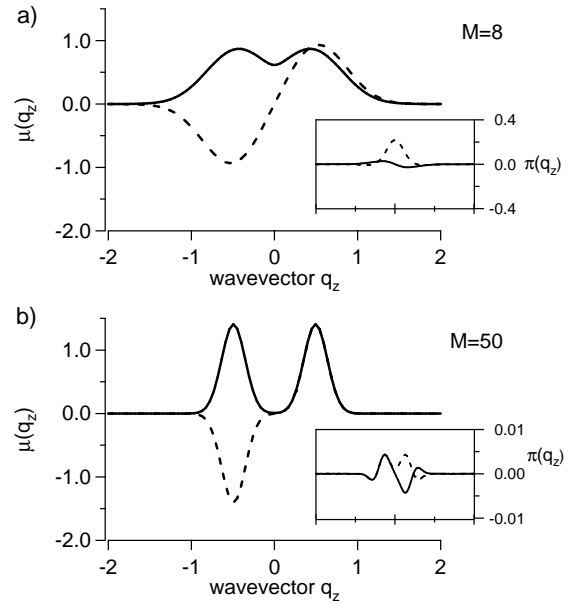}}
     \caption{The two lowest-energy eigenfunctions (solid line for the ground state, dashed for the excited state) which define
     the right- and left-well modes for the two-mode
     approximation.  Component wavefunctions in the lower ($\mu(q_z)$) and
     upper ($\pi(q_z)$, shown in insets) periodically-dressed
     states are shown for $M = 8$ and
     $M=50$.  As the spatial confinement is weakened (larger
     $M$), the wavefunctions become further localized in the minima of
     the double-well potential, and the population in the upper dressed state diminishes.
     Note the different parity of the $\mu$ and $\pi$ components, as
     discussed in the text.  Here $\hbar \Omega
     / E_k = 1/8$,  wavevectors are scaled by $k$, and the one-dimensional
     wavefunctions $\vec{\phi}_{0,1}(q_z)$ are shown.
     }
     \label{fig:wavefunctions}
    \end{center}
\end{figure}

These one-dimensional eigenstates, calculated using a restricted
basis set of Fourier components over the domain $-3 \leq q \leq
3$, are shown in Figure \ref{fig:wavefunctions} for two different
values of the mass parameter $M$, and for the condition of exact
Raman resonance $\delta = 0$ and Rabi frequency $\hbar \Omega /
E_k = 1/8$. For weak spatial confinement (large $M = 50$), the
lowest two energy eigenstates are indeed nearly entirely composed
of lower dressed states ($\pi(q_z)\simeq 0$) and are
well-approximated by the sum or difference of Gaussian right- and
left-well wavefunctions centered at the potential minima near
$q_z/k \simeq \pm 1/2$.  For stronger spatial confinement (smaller
$M = 8$), the enhanced ``momentum-space tunnelling'' causes the
wavefunctions to become less confined in the potential minima.
Concomitantly, a significant population appears in the upper
dressed states.

\begin{figure}
    \begin{center}
    \center{\includegraphics[width = 3 in]{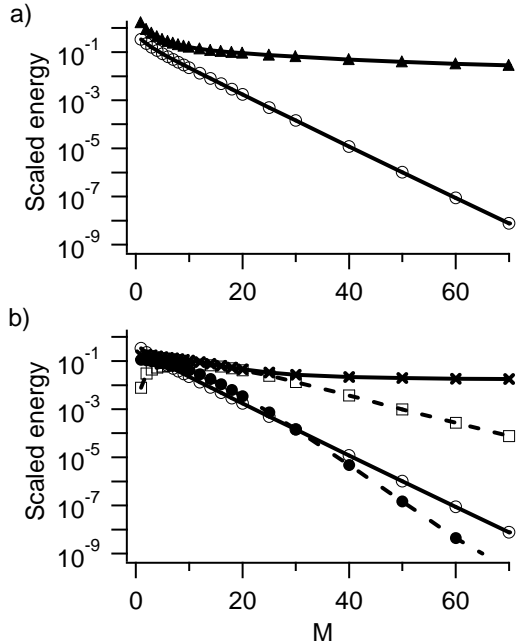}}
     \caption{Numerical calculations of tunnelling and interaction strengths.
     a) The tunnel splitting $J$ (open circles, solid line) becomes much smaller than the spacing between the second and third excited states
     (open triangles,solid line) for moderate values of $M$, establishing the validity of the two-mode approximation for weak confinement.
     b) Tunnelling energies $J$ (open circles, solid line), $J^{(1)}$ (open squares, dashed line) and $J^{(2)}$ (filled circles, dashed line)
     are suppressed for weaker confinement (exponentially with large $M$), while
     the strength of the momentum exchange energy $U$ (X's, solid line) remains large.  This provides a route to creating
     correlated many-body states adiabatically from uncorrelated states by gradually weakening the spatial confinement.  The energies in (a) are scaled
     by $E_k$.  Dimensionless interaction parameters are plotted as $-U / g \langle n \rangle$, $J^{(1)}/2 N g \langle n \rangle$,
     $J^{(2)} / (g \langle n \rangle / 2)$.  A Rabi frequency
     $\hbar \Omega / E_k = 1/8$ is chosen.
     }
     \label{fig:tunnel}
    \end{center}
\end{figure}

Parameters which enter into the two-mode Hamlitonian are derived
from the numerically-obtained wavefunctions, and are shown in
Figure \ref{fig:tunnel}. One finds the scaling  $J \sim \exp(-
M/4)$ as one expects \cite{merzfootnote}.  The numerical results
confirm the scaling behaviour for $J^{(1)}$ found in the weak
confinement (large $M$) limit (Eqs.\ \ref{eq:startscalings} --
\ref{eq:endscalings}), but differ slightly from the scaling
predicted for $J^{(2)}$ since the assumption that the right-well
state remains Gaussian in the vicinity of the left-well potential
minimum is incorrect. Nevertheless, $J^{(2)}$ is strongly
suppressed for large $M$, approximately as $\exp(-M/3)$. Also
shown is the energy splitting between the second and third lowest
eigenenergies; one sees that for weak spatial confinement the
lowest two energy eigenstates become well separated from the
remaining eigenspectrum, establishing the validity of the two-mode
approximation.

Finally, we stress that the two-mode treatment presented here
becomes invalid in the Thomas-Fermi regime, where the strength of
atomic interactions dominates the zero-point energy in the
confining potential.  In this situation, additional
single-particle states must be considered, resulting in a
complicated, self-consistent definition of the right- and
left-well states which depends on the number of atoms in these
states.  How to properly treat such a situation remains an open
question, and therefore an important subject for experimental
investigation. In related work, Montina and Arecchi \cite{mont03}
obtain right- and left-well states through use of the
Gross-Pitaevskii equation, but the validity of their treatment is
questionable for systems which are not completely
interaction-dominated.

\section{Application to Heisenberg-limited measurement \label{sec:heisenberg}}

Several theoretical works have pointed out the potential for
correlated many-body states for improving the precision of atomic
clocks \cite{boll96} and interferometers \cite{holl93,bouy97}, in
which phase shifts are measured at the Heisenberg limit $\Delta
\phi \propto 1/N$, rather than the standard quantum limit $\Delta
\phi \propto 1/\sqrt{N}$, where $N$ is the number of particles
used in a single run of the experiment.  In particular, Bollinger
\emph{et al.}\ described the use of the maximally-entangled state
in such a measurement \cite{boll96}.  In this section, we point
out how dynamical control over experimental parameters can be used
fruitfully to generate Schr\"{o}dinger-cat states suited for the
implementation of both Heisenberg-limited spectroscopy and
interferometry.

In the system we have described, repulsive interatomic
interactions lead to lowest energy many-body states which are the
even- and odd-superposition Schr\"{o}dinger cat states,
$|S_\pm\rangle = (|\Psi_R\rangle \pm |\Psi_L\rangle)/\sqrt{2}$.
The states $|\Psi_R\rangle$ and $|\Psi_L\rangle$ are
distinguishable many-body states associated with the right and
left potential wells, respectively, which tend toward the limiting
cases $|\Psi_R\rangle \rightarrow |0,N\rangle$ and $|\Psi_L\rangle
\rightarrow |N,0\rangle$ for strong interactions.  The energy
separation between these two states depends on the residual
overlap between $|\Psi_R\rangle$ and $|\Psi_L\rangle$ and the
number of atoms in the system (derived, e.g., in Ref.\
\cite{cira98super}). In the weak-confinement limit $M \gg 1$,
these right- and left-well states are comprised of the lower
dressed states, which are superpositions in both internal and
external degrees of freedom. Such a superposition appears to
complicate the use of these Schrodinger-cat states for measurement
applications.

However, after these Schrodinger-cat states are formed, the
dressed right- and left-well  states \emph{can be converted to
pure internal or external states adiabatically,} i.e.\ by
modification of experimental parameters on a time-scale which is
fast with respect to tunnelling times so as to preserve
entanglement, but slow with respect to timescales relevant to the
Raman coupling ($1/\Omega$). If the Rabi frequency $\Omega$ is
adiabatically lowered to zero (the Raman beams are slowly
extinguished), atoms in the right-well state would be
adiabatically converted to stationary trapped atoms in the
$|A\rangle$ internal state, while atoms in the left well would be
converted to stationary trapped atoms in the $|B\rangle$ state.
Thus would be prepared a state ideal for Heisenberg-limited
measurement of the internal state energy difference $\hbar
\omega_0$, potentially on a useful hyperfine clock transition.
Alternately, one could adiabatically ramp the detuning from the
Raman resonance to large positive or negative values.  For
example, an adiabatic downward sweep of the frequency difference
between the Raman laser beams ($\delta \rightarrow - \infty$)
lowers the energy of the $|A\rangle$ internal state with respect
to the $|B\rangle$ internal state in the rotating frame. This
adiabatically converts the right- and left-well states to
\emph{distinguishable momentum states} with identical internal
states (in this case the $|A\rangle$ state). One thereby prepares
a state ideal for Heisenberg-limited atomic interferometry.

Furthermore, the timed application of Raman coupling in this
system can be used as a ``magic beamsplitter'' \cite{lee02rosetta}
to prepare the Schr\"{o}dinger-cat state dynamically.  For
simplicity let us make the identification $|\Psi_R\rangle =
|0,N\rangle$ and $|\Psi_L\rangle = |N,0\rangle$.  Consider a
system prepared in the $|\Psi\rangle = |N, 0\rangle$ state with no
Raman coupling ($\Omega = 0$) -- this state corresponds to a
zero-temperature Bose-Einstein condensate of $N$ atoms in the
$|B\rangle$ internal state.  The Raman coupling is then turned on
adiabatically, as discussed above.  The initial state
$|\Psi\rangle$, being a superposition of the two-lowest energy
Schr\"{o}dinger-cat energy eigenstates, is now led to oscillate
coherently, and collectively, between the right- and left-well
states.  This oscillation is analogous to macroscopic quantum
tunnelling observed, for example, in the Rabi oscillations of
superconducting qubits \cite{chio03}.  If the Raman coupling is
left on for a duration which is $1/4$ of the Bohr period between
the even and odd Schr\"{o}dinger-cat states, the many-body state
$|\Psi\rangle$ will evolve as
\begin{equation}
|\Psi\rangle \rightarrow \frac{|S_+\rangle - i
|S_-\rangle}{\sqrt{2}} = \frac{1 - i}{2} |N,0\rangle + \frac{1 +
i}{2} |0, N\rangle
\end{equation}
The Raman beams can be then switched off -- with the
transformation of the state into a momentum-space or
internal-state-space Schr\"{o}dinger-cat state, as desired -- for
a duration $\tau$, during which a relative phase can accrue
between the two distinguishable portions of the
Schr\"{o}dinger-cat state, i.e.\
\begin{equation}
|\Psi\rangle \rightarrow \frac{1 - i}{2} |N,0\rangle + \frac{1 +
i}{2} e^{-i N \Delta \tau} |0, N\rangle
\end{equation}
Here $\Delta$ is the difference in energy between the
single-particle states to which the right- and left-well states
are connected when the Raman beams are extinguished.  For
instance, if we are following the implementation of a measurement
of the clock frequency between internal states, $\Delta = \delta$.

Finally, this state must be analyzed.   In the scheme of Bollinger
\emph{et al.}\ \cite{boll96}, a conventional $\pi/2$ pulse is
applied, and a precise measurement is made of the difference
between the number of atoms in the $|A\rangle$ state and the
$|B\rangle$ state -- one must establish whether this number
difference is even or odd, thus requiring single-atom precision in
the number counting.

Alternately, one may apply a \emph{second} ``magic beamsplitter''
pulse as before, thus preparing a state
\begin{equation}
|\Psi\rangle \rightarrow \frac{-i (1 - e^{-i N \Delta \tau})}{2}
|N,0\rangle + \frac{(1 + e^{-i N \Delta \tau})}{2} |0, N\rangle
\end{equation}
Measurements on this state would detect \emph{all atoms} in either
the right- or left-well states (say, internal states $|A\rangle$
or $|B\rangle$), with probabilities which vary periodically with
the free-evolution time $\tau$ with frequency $N \Delta$ which is
$N$ times higher than for a single-particle Ramsey-type
measurement. This is the source of the enhanced Heisenberg-level
sensitivity. This sensitivity persists even in the presence of
overall number fluctuations, presuming that one's goal is to
ascertain whether $\Delta = 0$.  However, the correct timing
required to yield a perfect ``magic beamsplitter'' will depend on
$N$ \cite{cira98super}.  Thus, attaining the highest precision
will still require a highly accurate determination of the total
atom number $N$, though perhaps not necessarily at the single-atom
level.  A further unexamined issue is how the finite temperature
of the atomic sample, which prevents the initial state
$|\Psi\rangle$ from being a pure Bose-Einstein condensate, will
impact the precision of these Schr\"{o}dinger-cat-based
measurements.

In conclusion, we have presented an analytic model which
identifies the many-body ground states of a weakly-interacting
Bose gas which is ``periodically dressed'' by continuous Raman
excitation and confined in an harmonic spatial potential.  The
system is analyzed in momentum space, wherein the balance between
tunnelling and weak interactions dictates whether the ground
states are uncorrelated product states of single-particle
wavefunctions, or highly correlated states.  It is found that
interactions in the physical system considered here have the
opposite effect as in the dual situation of a position-space
double-well potential, that is, repulsive interactions will lead
to Schr\"{o}dinger-cat states while attractive interactions will
lead to number-squeezed states with equal numbers of atoms in each
well.  Further, we show that experimental parameters can be used
to dynamically tune the interaction strength and tunnelling rates.
This degree of control can be used to generate maximally-entangled
states directly suitable for Heisenberg-limited metrology and
interferometry.

We thank Lorraine Sadler, Subhadeep Gupta, Joel Moore, and
Veronique Savalli for critical discussions and readings.  This
work was supported by the National Science Foundation under Grant
No.\ 0133999, by the Hellman Faculty Fund, the Sloan Foundation,
the Packard Foundation, and the University of California.

\bibliographystyle{apsrev}

\begin{thebibliography}{30}
\expandafter\ifx\csname
natexlab\endcsname\relax\def\natexlab#1{#1}\fi
\expandafter\ifx\csname bibnamefont\endcsname\relax
  \def\bibnamefont#1{#1}\fi
\expandafter\ifx\csname bibfnamefont\endcsname\relax
  \def\bibfnamefont#1{#1}\fi
\expandafter\ifx\csname citenamefont\endcsname\relax
  \def\citenamefont#1{#1}\fi
\expandafter\ifx\csname url\endcsname\relax
  \def\url#1{\texttt{#1}}\fi
\expandafter\ifx\csname
urlprefix\endcsname\relax\def\urlprefix{URL }\fi
\providecommand{\bibinfo}[2]{#2}
\providecommand{\eprint}[2][]{\url{#2}}

\bibitem[{\citenamefont{Bennett et~al.}(1993)}]{benn93tele}
\bibinfo{author}{\bibfnamefont{C.~H.} \bibnamefont{Bennett}}
  \bibnamefont{et~al.}, \bibinfo{journal}{Physical Review Letters}
  \textbf{\bibinfo{volume}{70}}, \bibinfo{pages}{1895} (\bibinfo{year}{1993}).

\bibitem[{\citenamefont{Caves}(1981)}]{cave81}
\bibinfo{author}{\bibfnamefont{C.~M.} \bibnamefont{Caves}},
  \bibinfo{journal}{Physical Review D} \textbf{\bibinfo{volume}{23}},
  \bibinfo{pages}{1693} (\bibinfo{year}{1981}).

\bibitem[{\citenamefont{Wineland et~al.}(1992)}]{wine92squeeze}
\bibinfo{author}{\bibfnamefont{D.~J.} \bibnamefont{Wineland}}
  \bibnamefont{et~al.}, \bibinfo{journal}{Physical Review A}
  \textbf{\bibinfo{volume}{46}}, \bibinfo{pages}{R6797} (\bibinfo{year}{1992}).

\bibitem[{\citenamefont{Bollinger et~al.}(1996)}]{boll96}
\bibinfo{author}{\bibfnamefont{J.~J.} \bibnamefont{Bollinger}}
  \bibnamefont{et~al.}, \bibinfo{journal}{Physical Review A}
  \textbf{\bibinfo{volume}{54}}, \bibinfo{pages}{R4649} (\bibinfo{year}{1996}).

\bibitem[{\citenamefont{Bouyer and Kasevich}(1997)}]{bouy97}
\bibinfo{author}{\bibfnamefont{P.}~\bibnamefont{Bouyer}} \bibnamefont{and}
  \bibinfo{author}{\bibfnamefont{M.~A.} \bibnamefont{Kasevich}},
  \bibinfo{journal}{Physical Review A} \textbf{\bibinfo{volume}{56}},
  \bibinfo{pages}{R1083} (\bibinfo{year}{1997}).

\bibitem[{\citenamefont{Julsgaard et~al.}(2001)}]{juls01}
\bibinfo{author}{\bibfnamefont{B.}~\bibnamefont{Julsgaard}}
  \bibnamefont{et~al.}, \bibinfo{journal}{Nature}
  \textbf{\bibinfo{volume}{413}}, \bibinfo{pages}{400} (\bibinfo{year}{2001}).

\bibitem[{\citenamefont{Sackett et~al.}(2000)}]{sack00four}
\bibinfo{author}{\bibfnamefont{C.~A.} \bibnamefont{Sackett}}
  \bibnamefont{et~al.}, \bibinfo{journal}{Nature}
  \textbf{\bibinfo{volume}{404}}, \bibinfo{pages}{256} (\bibinfo{year}{2000}).

\bibitem[{\citenamefont{Brune et~al.}(1996)}]{brun96meter}
\bibinfo{author}{\bibfnamefont{M.}~\bibnamefont{Brune}} \bibnamefont{et~al.},
  \bibinfo{journal}{Physical Review Letters} \textbf{\bibinfo{volume}{77}},
  \bibinfo{pages}{4887} (\bibinfo{year}{1996}).

\bibitem[{\citenamefont{Pashkin et~al.}(2003)}]{pash03}
\bibinfo{author}{\bibfnamefont{Y.~A.} \bibnamefont{Pashkin}}
  \bibnamefont{et~al.}, \bibinfo{journal}{Nature}
  \textbf{\bibinfo{volume}{421}}, \bibinfo{pages}{823} (\bibinfo{year}{2003}).

\bibitem[{\citenamefont{Berkley et~al.}(2003)}]{well03}
\bibinfo{author}{\bibfnamefont{A.~J.} \bibnamefont{Berkley}}
  \bibnamefont{et~al.}, \bibinfo{journal}{Science}
  \textbf{\bibinfo{volume}{300}}, \bibinfo{pages}{1548} (\bibinfo{year}{2003}).

\bibitem[{\citenamefont{Mandel et~al.}()}]{mand03cont}
\bibinfo{author}{\bibfnamefont{O.}~\bibnamefont{Mandel}} \bibnamefont{et~al.},
  \bibinfo{note}{preprint ArXiv:quant-ph/0308080}.

\bibitem[{\citenamefont{Andr\'{e} and Lukin}(2002)}]{andr02corr}
\bibinfo{author}{\bibfnamefont{A.}~\bibnamefont{Andr\'{e}}} \bibnamefont{and}
  \bibinfo{author}{\bibfnamefont{M.~D.} \bibnamefont{Lukin}},
  \bibinfo{journal}{Physical Review A} \textbf{\bibinfo{volume}{65}},
  \bibinfo{pages}{053819} (\bibinfo{year}{2002}).

\bibitem[{\citenamefont{Sorensen and Molmer}(2002)}]{sore02cavity}
\bibinfo{author}{\bibfnamefont{A.~S.} \bibnamefont{Sorensen}} \bibnamefont{and}
  \bibinfo{author}{\bibfnamefont{K.}~\bibnamefont{Molmer}},
  \bibinfo{journal}{Physical Review A} \textbf{\bibinfo{volume}{66}},
  \bibinfo{pages}{022314} (\bibinfo{year}{2002}).

\bibitem[{\citenamefont{Imamoglu et~al.}(1997)}]{imam97}
\bibinfo{author}{\bibfnamefont{A.}~\bibnamefont{Imamoglu}}
  \bibnamefont{et~al.}, \bibinfo{journal}{Physical Review Letters}
  \textbf{\bibinfo{volume}{78}}, \bibinfo{pages}{2511} (\bibinfo{year}{1997}).

\bibitem[{\citenamefont{Cirac et~al.}(1998)}]{cira98super}
\bibinfo{author}{\bibfnamefont{J.~I.} \bibnamefont{Cirac}}
  \bibnamefont{et~al.}, \bibinfo{journal}{Physical Review A}
  \textbf{\bibinfo{volume}{57}}, \bibinfo{pages}{1208} (\bibinfo{year}{1998}).

\bibitem[{\citenamefont{Javanainen and Ivanov}(1999)}]{java99split}
\bibinfo{author}{\bibfnamefont{J.}~\bibnamefont{Javanainen}} \bibnamefont{and}
  \bibinfo{author}{\bibfnamefont{M.~Y.} \bibnamefont{Ivanov}},
  \bibinfo{journal}{Physical Review A} \textbf{\bibinfo{volume}{60}},
  \bibinfo{pages}{2351} (\bibinfo{year}{1999}).

\bibitem[{\citenamefont{Spekkens and Sipe}(1999)}]{spek99}
\bibinfo{author}{\bibfnamefont{R.~W.} \bibnamefont{Spekkens}} \bibnamefont{and}
  \bibinfo{author}{\bibfnamefont{J.~E.} \bibnamefont{Sipe}},
  \bibinfo{journal}{Physical Review A} \textbf{\bibinfo{volume}{59}},
  \bibinfo{pages}{3868} (\bibinfo{year}{1999}).

\bibitem[{\citenamefont{Orzel et~al.}(2001)}]{orze01squeeze}
\bibinfo{author}{\bibfnamefont{C.}~\bibnamefont{Orzel}} \bibnamefont{et~al.},
  \bibinfo{journal}{Science} \textbf{\bibinfo{volume}{291}},
  \bibinfo{pages}{2386} (\bibinfo{year}{2001}).

\bibitem[{\citenamefont{Greiner et~al.}(2002)}]{grei02mott}
\bibinfo{author}{\bibfnamefont{M.}~\bibnamefont{Greiner}} \bibnamefont{et~al.},
  \bibinfo{journal}{Nature} \textbf{\bibinfo{volume}{415}}, \bibinfo{pages}{39}
  (\bibinfo{year}{2002}).

\bibitem[{\citenamefont{Sackett et~al.}(1999)}]{sack99collapse}
\bibinfo{author}{\bibfnamefont{C.~A.} \bibnamefont{Sackett}}
  \bibnamefont{et~al.}, \bibinfo{journal}{Physical Review Letters}
  \textbf{\bibinfo{volume}{82}}, \bibinfo{pages}{876} (\bibinfo{year}{1999}).

\bibitem[{\citenamefont{Higbie and Stamper-Kurn}(2002)}]{higb02}
\bibinfo{author}{\bibfnamefont{J.}~\bibnamefont{Higbie}} \bibnamefont{and}
  \bibinfo{author}{\bibfnamefont{D.~M.} \bibnamefont{Stamper-Kurn}},
  \bibinfo{journal}{Physical Review Letters} \textbf{\bibinfo{volume}{88}},
  \bibinfo{pages}{090401} (\bibinfo{year}{2002}).

\bibitem[{\citenamefont{Stamper-Kurn}(2003)}]{stam03anis}
\bibinfo{author}{\bibfnamefont{D.~M.} \bibnamefont{Stamper-Kurn}},
  \bibinfo{journal}{New Journal of Physics} \textbf{\bibinfo{volume}{5}},
  \bibinfo{pages}{50} (\bibinfo{year}{2003}).

\bibitem[{\citenamefont{Montina and Arecchi}(2003)}]{mont03}
\bibinfo{author}{\bibfnamefont{A.}~\bibnamefont{Montina}} \bibnamefont{and}
  \bibinfo{author}{\bibfnamefont{F.~T.} \bibnamefont{Arecchi}},
  \bibinfo{journal}{Physical Review A} \textbf{\bibinfo{volume}{67}},
  \bibinfo{pages}{23616} (\bibinfo{year}{2003}).

\bibitem[{\citenamefont{Leggett}(2001)}]{legg00rmp}
\bibinfo{author}{\bibfnamefont{A.~J.} \bibnamefont{Leggett}},
  \bibinfo{journal}{Reviews of Modern Physics} \textbf{\bibinfo{volume}{73}},
  \bibinfo{pages}{307} (\bibinfo{year}{2001}).

\bibitem[{rad()}]{radial_symmetry_footnote}
\bibinfo{note}{Such a potential for ultracold alkali atoms might be created by
  optical trapping at large detunings and/or with linearly polarized light, or
  by magnetic trapping of two states with identical magnetic moments. For
  simplicity, we have also chosen a radially-symmetric trap, though our
  analysis is only slightly modified by trap asymmetry as long as the Raman
  momentum transfer $\hbar {\bf{k}}$ is aligned along an axis of the trap.}

\bibitem[{sym()}]{symmetry_footnote}
\bibinfo{note}{Defining the operator $\mathcal{P}^* = \mathcal{P} \sigma_z$
  where $\mathcal{P}$ is the parity operator and $\sigma_z$ the Pauli matrix,
  we find for $\delta = 0$ that $[\mathcal{P}^*, \hami] = 0$. Given
  $\phi^{PD}_{0,1}$ are real, we obtain the relations $\mathcal{N}_{R
  R}({\bf{q}}) = \mathcal{N}_{L L}({\bf{q}}) = \mathcal{N}_{R R}(-{\bf{q}})$,
  and $\mathcal{N}_{R L}({\bf{q}}) = \mathcal{N}_{L R}(-{\bf{q}})$.}

\bibitem[{mer()}]{merzfootnote}
\bibinfo{note}{A similar scalar one-dimensional double-well potential -- two
  offset parabolae connected with a discontinuous slope -- is treated
  analytically as a textbook problem. See for example E. Merzbacher,
  \emph{Quantum Mechanics}, 2nd Ed.\ (John Wiley and Sons, New York, 1970),
  Eq.\ 5.61.}

\bibitem[{\citenamefont{Holland and Burnett}(1993)}]{holl93}
\bibinfo{author}{\bibfnamefont{M.~J.} \bibnamefont{Holland}} \bibnamefont{and}
  \bibinfo{author}{\bibfnamefont{K.}~\bibnamefont{Burnett}},
  \bibinfo{journal}{Physical Review Letters} \textbf{\bibinfo{volume}{71}},
  \bibinfo{pages}{1355} (\bibinfo{year}{1993}).

\bibitem[{\citenamefont{Lee et~al.}(2002)}]{lee02rosetta}
\bibinfo{author}{\bibfnamefont{H.}~\bibnamefont{Lee}} \bibnamefont{et~al.},
  \bibinfo{journal}{Journal of Modern Optics} \textbf{\bibinfo{volume}{49}},
  \bibinfo{pages}{2325} (\bibinfo{year}{2002}).

\bibitem[{\citenamefont{Chiorescu et~al.}(2003)}]{chio03}
\bibinfo{author}{\bibfnamefont{I.}~\bibnamefont{Chiorescu}}
  \bibnamefont{et~al.}, \bibinfo{journal}{Science}
  \textbf{\bibinfo{volume}{299}}, \bibinfo{pages}{1868} (\bibinfo{year}{2003}).

\end{thebibliography}

\end{document}